\documentclass[12pt]{iopart}

\usepackage{epsfig}
\usepackage{epstopdf}

\begin{document}

\title[Finite-size scaling of the stochastic 
susceptible-infected-recovered model]
{Finite-size scaling of the stochastic 
susceptible-infected-recovered model}

\author{David R. de Souza$^{\,1}$, T\^ania Tom\'e$^{\,1}$ 
and Robert M. Ziff$^{\,2}$}

\address{$^1$Instituto de F\'{\i}sica,
Universidade de S\~{a}o Paulo, \\
Caixa Postal 66318,
05314-970 S\~{a}o Paulo, S\~{a}o Paulo, Brazil \\
$^2$Michigan Center for Theoretical Physics \\
and Department of Chemical Engineering, University of Michigan \\
Ann Arbor, MI 48109-2136, USA }

\begin{abstract}

The critical behavior of the stochastic
susceptible-infected-recovered model on a square lattice
is obtained by numerical simulations and 
finite-size scaling. The order parameter as well as the 
distribution in the number of recovered individuals is
determined as a function of the infection rate for 
several values of the system size. 
The analysis around criticality is 
obtained by exploring the close relationship between the 
present model and standard percolation theory. 
The quantity $UP$, equal to the ratio $U$ between 
the second moment and the squared first
moment of the size distribution multiplied by 
the order parameter $P$, is shown 
to have, for a square system, a universal value
$1.0167(1)$ that is the same as for site and 
bond percolation, confirming further that the
SIR model is also in the percolation class.

\end{abstract}


\maketitle

\section{Introduction}

The spread of an epidemic among a community of individuals
has been described by several types of models,
either deterministic or stochastic
\cite{kermack27,bailey53,bailey57,mollison77,renshaw91,hastings96,keeling08}.
Among the latter we find models in which the
space structure is explicitly taken into account
\cite{grassberger83,cardy85,satulovsky94,satulovsky97,antal01,
dammer03,dammer04,arashiro07,schutz08,assis09,desouza10,tome10,tome10a}. 
These models
are defined on a lattice which represents the space where the community
lives, and in which each site is occupied by just one individual.
In the susceptible-infected-recovered (SIR) stochastic lattice model 
\cite{grassberger83,cardy85,antal01,
dammer03,dammer04,arashiro07,schutz08,assis09,desouza10,tome10,tome10a},
each individual can be either susceptible (S), infected (I)
or recovered (R). A susceptible individual becomes infected
(S$\to$I) through an autocatalytic reaction, and an infected individual
recovers (I$\to$R) spontaneously. This model
describes an epidemic in which the immunization is 
permanent; in other words, once an individual recovers,
it becomes immune forever.

The main features of the SIR model are as follows.
When the rate of infection of a susceptible individual by
an infected individual is small compared to the immunization rate, 
there is no spreading of the disease.
Increasing the infection rate, one reaches a critical
value above which the infection spreads over the whole lattice.
The transition from one regime to the other is regarded
a continuous phase transition whose critical behavior places
the model into the dynamic percolation universality class,
which corresponds to the standard percolation class with the 
addition of dynamical growth exponents
\cite{grassberger83,cardy85}.  
As shown below, the SIR model can be described by
just one parameter, either the reduced infection rate $b$
or the reduced immunization rate $c=1-b$.
The phase transition occurs at a critical value 
$c=c_c$, which has been estimated as $0.1765(5)$ on a square lattice
by means of time-dependent numerical
simulations \cite{desouza10}. A more accurate result
$c_c=0.1765005(10)$
 was later
determined by extensive numerical simulation through the use
of a technique borrowed from standard percolation theory \cite{tome10}. 

A close relationship exists between the SIR model
and dynamic (isotropic) percolation 
\cite{mollison77,grassberger83,tome10,tome10a,kuulasmaa82,newman02}. 
Starting from a single infected
individual in a lattice full of susceptibles,
a cluster composed of infected and recovered
individuals grows, the infected individuals staying at
the border of the cluster and the recovered individuals inside it.
Eventually the cluster becomes composed of recovered individuals only
due to the spontaneous immunization. 
For $c<c_c$ (the spreading regime), an infinite cluster of
recovered individuals percolates the
whole lattice. For $c>c_c$ (the non-spreading regime),
only finite clusters are present. It has been shown \cite{tome10}
that the cluster probability distribution obeys
the same scaling laws as apply to standard percolation models.

Here we report numerical simulations and finite-size scaling
analysis to obtain the critical behavior of the stochastic
SIR model on a square lattice.
To this end we determine the order parameter $P$
(defined below),
the mean number of recovered sites $S$, and the mean
value of the squared number of recovered sites $M$. These
quantities are determined as a function of $c$ and the linear
system size $L$. We show that the ratio $U=M/S^2$ 
between the second moment and the square of the first moment 
is not independent of the system size $L$, at criticality,
as occurs, e.\ g., in the contact model \cite{dickman05,tome05}. 
Instead, we show here that the quantity that
is independent of the system size at criticality,
and therefore universal, is 
the product $UP$ between this ratio and the order parameter,
consistent with the dynamical percolation universality class.

This paper is organized as follows. The stochastic SIR model
on a regular lattice is defined in section \ref{2} together with
numerical simulations.
In section \ref{3} we explain how the model is related to percolation
around criticality.
The finite-size analysis is introduced in section \ref{4}.
A conclusion is drawn in the last section.

\section{Definition and simulations}
\label{2}

The stochastic SIR model is defined on a regular lattice 
of $N$ sites as follows. At each time step a site is chosen at 
random and the time is incremented by an amount equal to $1/N$.
If the chosen site is in state S then it becomes I with probability 
$b$ multiplied by the fraction of nearest-neighbor sites in state I. 
If the chosen site 
is in state I then it becomes R with probability $c=1-b$. 
If it is in state R it remains in this state. 
The number of individuals of type S, I and R are denoted by
$N_\mathrm S$, $N_\mathrm I$ and $N_\mathrm R$. The total number of individuals
equals the total number of sites of the lattice,
$N_\mathrm S+ N_\mathrm I + N_\mathrm R = N$.

The quantities that we have measured in the Monte Carlo
simulation are the following: the mean number of recovered
individuals 
\begin{equation}
S = \langle N_\mathrm R \rangle,
\end{equation}
the mean value of the square of the number of recovered individuals
\begin{equation}
M = \langle N_\mathrm R^2 \rangle,
\end{equation}
and the order parameter $P$, defined below.
We also considered
the ratio $U$ between the second moment $M$ and the square
of the first moment $S$ of the probability distribution of 
recovered individuals, that is,
\begin{equation}
U = \frac{M}{S^2}.
\end{equation}

The simulations were performed on a square lattice of $N=L^2$
sites and periodic boundary conditions.
We begin with an infected individual placed at the
center of the lattice full of susceptible individuals. 
To speed up the simulations we keep a list of the I sites.
At each step of the simulation we choose randomly an I site 
among the list of the $N_\mathrm I$ I sites.  (If we were interested
in the time, which is not the case here, we would increment it by an amount 
equal to $1/N_\mathrm I$.)  With probability $c$ the chosen I site becomes
an R site. With the complementary probability $b=1-c$
we choose one of its four nearest-neighbor sites;
if the nearest-neighbor site is an S site, it becomes an I site, otherwise
it remains unchanged. These rules are equivalent to the definition 
of the model given in the beginning of this section.

\begin{figure}
\centering
\epsfig{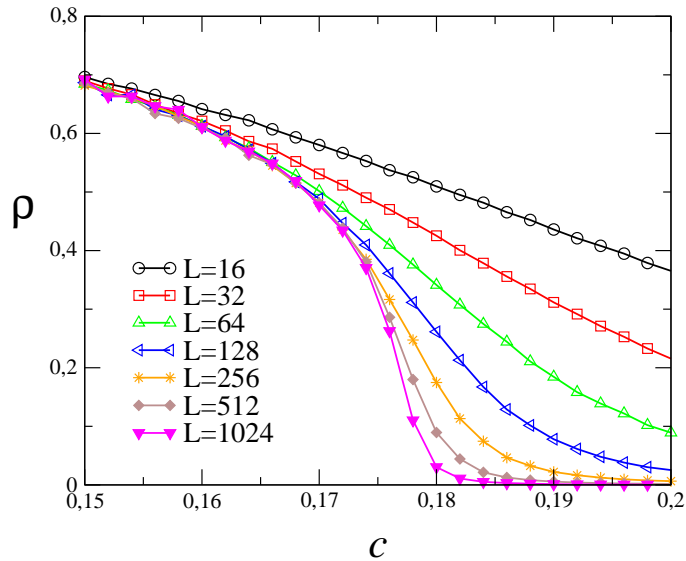}
\caption{Density of recovered individuals $\rho=S/N$ 
versus $c$ for several values of the linear system size $L$.}
\label{dens_recxc}
\end{figure}

\begin{figure}
\centering
\epsfig{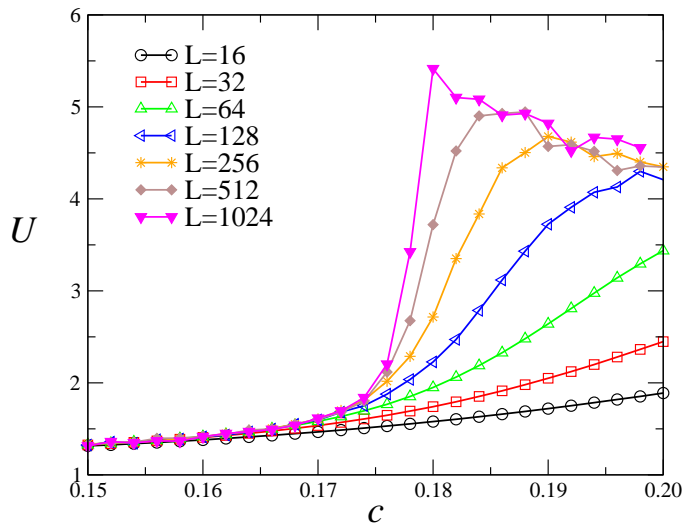}
\caption{Ratio $U=M/S^2$ versus $c$ for several values
of the linear system size $L$.}
\label{cumxc}
\end{figure}

Here we are interested only in the stationary states, 
which are characterized by the absence of infected sites. 
Starting from a single infected site, the number
of infected sites may increase but eventually decreases
and vanishes. Without infected sites there is no activity
and the dynamics stops. The stationary state is then an absorbing
state constituted of S and R sites only. For each
value of the parameter $c$ and linear size $L$ we performed a set of 
independent runs, ranging from $10^7$ to $10^8$,
and measured the quantities $S$, $M$ and $P$
related to the final clusters of R sites.
In figure \ref{dens_recxc} we show the density of
recovered individuals $\rho=S/N$ as a function of $c$ for
several values of the system size $L$.
In figure \ref{cumxc} we show the ratio $U=M/S^2$
as a function of $c$ for several values of the system
size $L$.

\begin{figure}
\centering
\epsfig{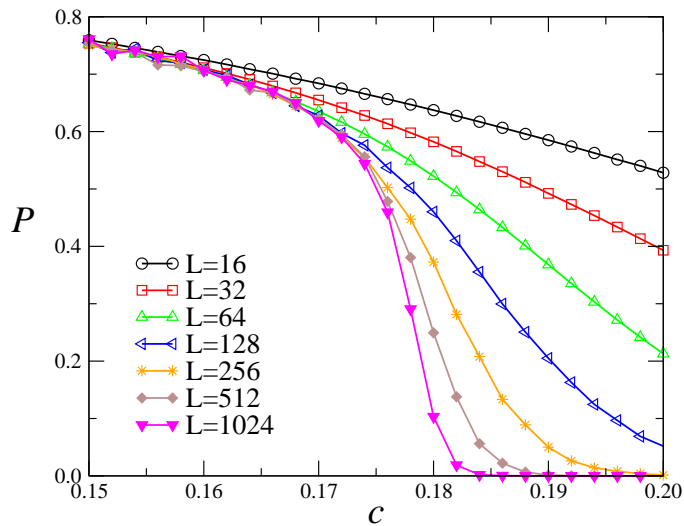}
\caption{Order parameter $P$ versus $c$ for several values
of the linear system size $L$.}
\label{pxc}
\end{figure}

To find the order parameter $P$, 
we checked whether the growing cluster of infected individuals
reached the border of the lattice, and 
the fraction of runs for which this happens is defined as $P$. 
In the thermodynamic limit, it becomes the probability that the
central site belongs to the infinite cluster
in accordance with the definition of order parameter
in standard percolation theory.  
In figure \ref{pxc} we show $P$
versus $c$ for several values of the linear size $L$.

We note that in percolation one often uses the
fraction of sites belonging to the largest cluster in a lattice fully
occupied with clusters as the order parameter.
However, this method cannot be used here because by definition there is
only one cluster in each epidemic sample.   Note also that
in our definition of $P$, one can think of the system as
having open boundary conditions and we are finding
if the cluster hits that boundary; this is 
related of the idea
of midpoint percolation considered recently \cite{baek10}.
An equivalent interpretation is that we have an infinite lattice, and we 
are seeing if the epidemic starting from the origin crosses the 
$L \times L$ boundary.

\section{Relation to percolation}
\label{3}

Here we summarize some results of the percolation theory
\cite{stauffer94,lorenz98,newman01,lee08,feng08}
that will be useful for showing the relation to the SIR model.
In standard percolation theory the probability that a site belongs to a cluster
of size $s$ is 
\begin{equation}
P_s=s n_s,
\end{equation}
where $n_s$ is the mean number of clusters of size $s$ per lattice site. 
(For site percolation, one should also technically multiply this by $p$, the
probability that the site is occupied, but we will suppress this factor.)
From this quantity one obtains the mean epidemic size $S$,
\begin{equation}
S = \sum_{s}sP_{s} = \sum_{s}s^2 n_{s},  
\end{equation}
and the mean-square epidemic cluster size $M$, 
\begin{equation}
M =\sum_{s} s^2 P_s=\sum_{s}s^3 n_{s}.
\end{equation}
The order parameter $P$ is the probability that a
site belongs to the infinite percolating cluster.

Around the critical point $p=p_c$ for an infinite system, 
these quantities behave as
\begin{equation}
P \sim \varepsilon^\beta   \qquad \varepsilon \ge  0,
\label{8}
\end{equation}
\begin{equation}
S \sim |\varepsilon|^{-\gamma },
\label{9}  
\end{equation}
and
\begin{equation}
M \sim |\varepsilon|^{-\beta-2\gamma},
\label{10}
\end{equation}
where $\varepsilon = p-p_{c}$ 
and $\beta$ and $\gamma$ are critical exponents 
associated to the order parameter 
and to the mean cluster size, respectively.
The quantity $p$ is the parameter associated with percolation 
problem and $p_c$ is its critical value.
In site (bond) percolation the parameter $p$ is 
the probability that a site (bond) is occupied.

It has been argued that the clusters of recovered individuals
generated by the dynamics of the stochastic SIR model
follow the statistics of the cluster size distribution
of the standard percolation theory summarized here \cite{grassberger83}.
The mean cluster size in percolation theory is then identified
as the mean number of recovered individuals.
This leads us to identify the quantities $P$, $S$ and $M$ defined
in the previous section for the SIR model
with the quantities $P$, $S$ and $M$ defined in this 
section. As a consequence, their critical behavior is given 
equations (\ref{8}), (\ref{9}) and (\ref{10}) with
\begin{equation}
\varepsilon=c - c_c.
\end{equation}
An implication of the critical behavior (\ref{9}) and (\ref{10})
of $S$ and $M$ is that the ratio
$U=M/S^2$ should behave as
\begin{equation}
U \sim |\varepsilon| ^{-\beta}.
\end{equation}

\begin{figure}
\centering
\epsfig{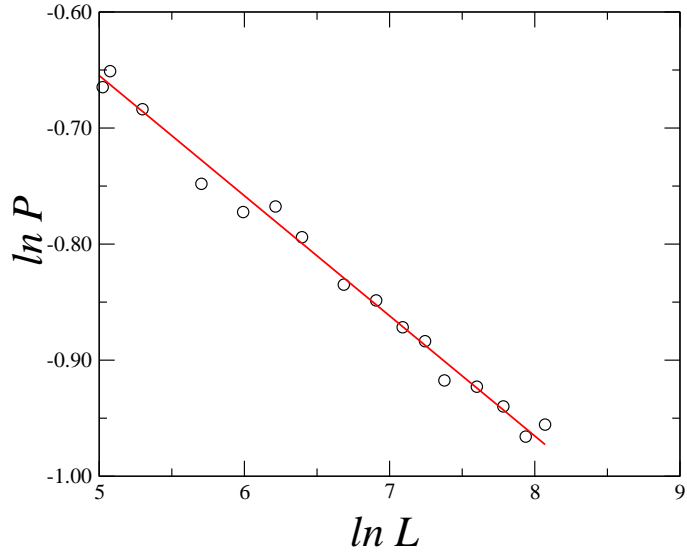}
\caption{Log-log plot of the order parameter $P$ versus $L$ 
for  $c=0.1765$. The slope
of the data points gives the value 
$\beta/\nu_\perp=0.1048$.}
\label{pbordaxL_rq}
\end{figure}

\begin{figure}
\centering
\epsfig{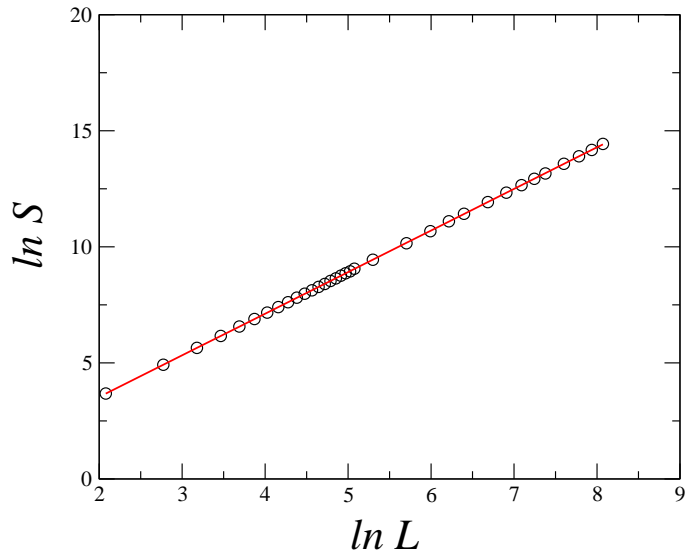}
\caption{Log-log plot of the mean number of recovered $S$ 
versus $L$ for $c=0.1765$.
The slope of the data points  gives the value 
$\gamma/\nu_\perp=1.7923$.}
\label{NzxL_rq}
\end{figure}

\section{Finite-size scaling}
\label{4}

We start from the assumption that a standard finite-size scaling analysis can 
be performed to obtain the critical behavior of this model. We assume that the
phase transition in the SIR model is characterized by a correlation length 
$\xi$ which diverges in the limit where the system is infinite as 
\begin{equation}
\xi \sim |\varepsilon|^{-\nu_\perp}.
\end{equation}
Here we use $\nu_\perp$ rather than the usual $\nu$ of standard 
percolation because
we consider the SIR cluster growth as a dynamical process.
The linear size of the
system $L$ scales as $\xi$ and a finite-system quantity 
$A_{L}$ will behave according to the finite-size scaling as
\begin{equation}
A_{L}=L^{\theta/{\nu_\perp }}\hat{A}(L^{1/{\nu_\perp }} \varepsilon),
\end{equation}
where $\hat{A}(X)$ is a universal function. 
The exponent $\theta$ describes the behavior of $A_L$ 
in the limit of the infinite system, that is,
\begin{equation}
A_{\infty } \sim |\varepsilon|^{-\theta }.
\end{equation}
Using the finite-size scaling, we may write the following
relations for the quantities $P$, $S$, $M$ and $U$,
\begin{equation}
P=L^{-\beta/{\nu_\perp }}\hat{P}(L^{1/{\nu_\perp }} \varepsilon),
\label{Pfinite}
\end{equation}
\begin{equation}
S=L^{\gamma/{\nu_\perp }}\hat{S}(L^{1/{\nu_\perp }} \varepsilon),
\label{Sfinite}
\end{equation}
\begin{equation}
M=L^{(\beta+2\gamma)/{\nu_\perp }}\hat{M}(L^{1/{\nu_\perp }} \varepsilon),
\end{equation}
\begin{equation}
U=L^{\beta/{\nu_\perp }}\hat{U}(L^{1/{\nu_\perp }} \varepsilon),
\end{equation}
and at the critical point $\varepsilon=0$, we have
\begin{equation}
P \sim L^{-\beta/{\nu_\perp }},
\end{equation} 
\begin{equation}
S \sim L^{\gamma/{\nu_\perp }},
\end{equation}
\begin{equation}
U \sim L^{\beta/{\nu_\perp }}.
\label{Ufinite}
\end{equation}
The quantities $P$ and $S$ are plotted as a function
of $L$ in figures \ref{pbordaxL_rq},
and \ref{NzxL_rq}. Each curve was obtained by performing
a number of runs of the order of $10^7$.
From the log-log plots we may estimate the critical exponents.
From the slope of a straight line 
fitted to the data points of figure \ref{pbordaxL_rq} we 
get the value $\beta/\nu_\perp=0.1048$ and 
from figure \ref{NzxL_rq} we get
the exponent $\gamma/\nu_\perp=1.7923$.
These results should be compared with the exact results
$\beta/\nu_\perp=5/48=0.1042$ and 
$\gamma/\nu_\perp=43/24=1.792$
coming from the exact
values $\beta=5/36$, $\gamma=43/18$ and $\nu_\perp=4/3$ 
of percolation in two dimensions.

\begin{figure}
\centering
\epsfig{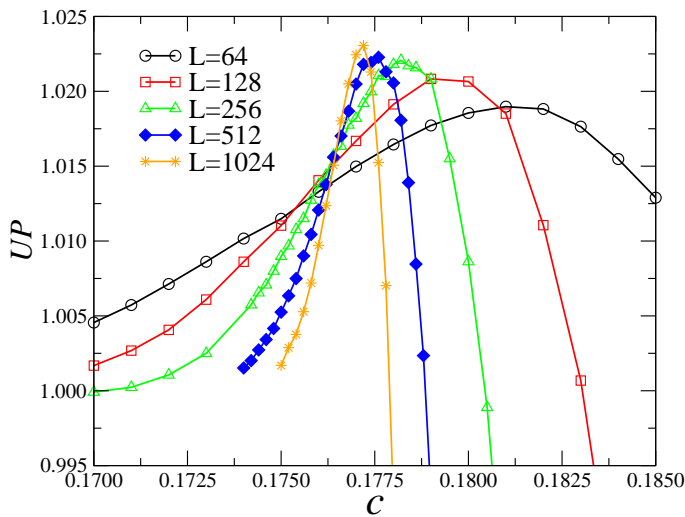}
\caption{Product $UP$ as a function of $c$ for several values
of the size $L$.}
\label{upxp}
\end{figure}

If we multiply equations (\ref{Pfinite}) and (\ref{Ufinite}) 
for the order parameter $P$ and the ratio $U$ we get
\begin{equation}
UP=\hat{F}(L^{1/{\nu_\perp }} \varepsilon).
\end{equation}
At the critical point $\varepsilon=0$ the product $UP$ 
is then a quantity independent of $L$ and
may be used to located the critical point.
Figure \ref{upxp} shows a plot of the quantity $UP$
versus $c$ for several values of the system size $L$.
We see that the curves indeed cross each other,
for sufficiently large values of $L$, at a point
identified as the critical point. From the plot we get
the value $c_c=0.17650(2)$ in agreement with the 
result $c_c=0.1765005(10)$ \cite{tome10}, and the
value $1.0167(1)$ for $UP$ at the critical point.

\begin{figure}
\centering
\epsfig{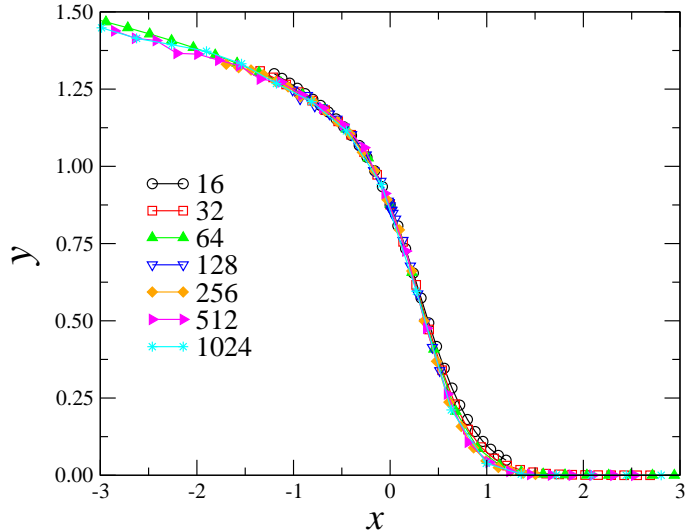}
\caption{Data collapse of the order parameter $P$ versus $c$ 
and $L$. The quantities $y$ and $x$ 
are defined by $y=PL^{\beta/\nu_\perp}$ and 
$x=\varepsilon L^{1/\nu_\perp}$ where $\varepsilon=c-c_c$.
The critical values used are $\beta/\nu_\perp=0.1048$, 
$\nu_\perp=1.333$ and $c=0.1765$.}
\label{pb_reescalado}
\end{figure}

\begin{figure}
\centering
\epsfig{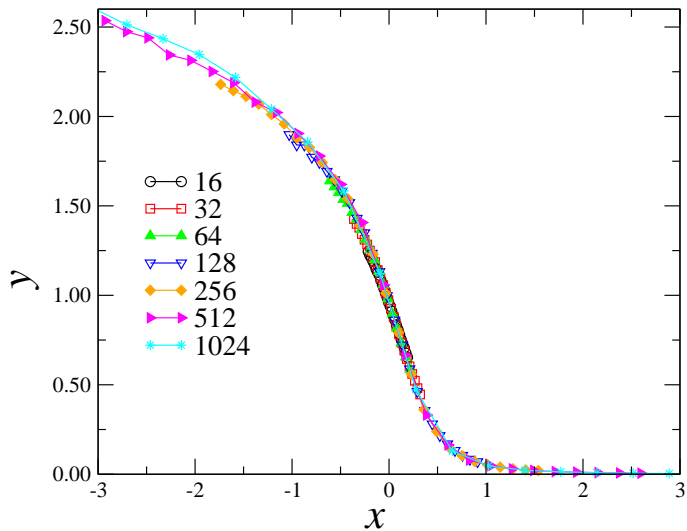}
\caption{Data collapse of the mean number of recovered $S$ versus $c$ 
and $L$. The quantities $y$ and $x$ 
are defined by $y=S L^{-\gamma/\nu_\perp}$ and 
$x=\varepsilon L^{1/\nu_\perp}$ where $\varepsilon=c-c_c$.
The critical values used are $\gamma/\nu_\perp=1.792$ and
$\nu_\perp=1.333$ and $c=0.1765$.}
\label{nz_reescalado}
\end{figure}

\begin{figure}
\centering
\epsfig{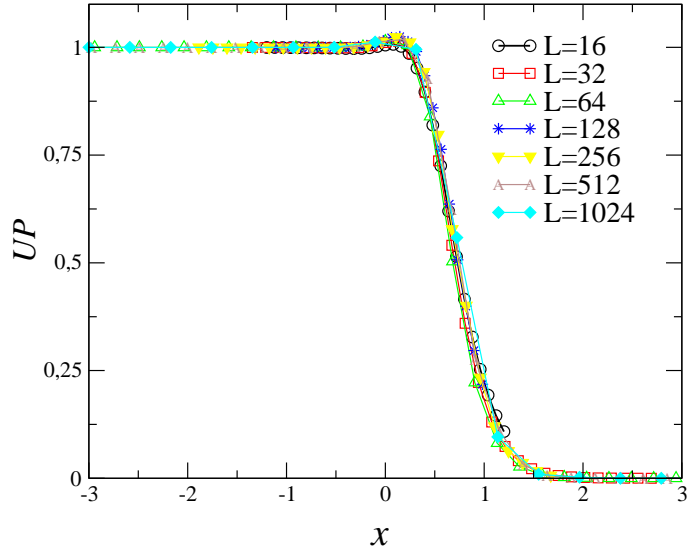}
\caption{Data collapse of the ratio $UP$ versus $c$ 
and $L$. The quantity $x$ is defined by
$x=\varepsilon L^{1/\nu_\perp}$ where $\varepsilon=c-c_c$.
The critical values used are $\nu_\perp=1.333$ and $c_c=0.1765$.}
\label{up_colapso}
\end{figure}

Using the critical points and the critical exponents
we have done a data collapse for the quantities
$P$, $S$ and $UP$, shown in figures
\ref{pb_reescalado}, \ref{nz_reescalado} and 
\ref{up_colapso}, respectively. These plots 
confirm that the critical behavior of the stochastic
SIR model obeys the finite-size scaling 
defined above.

\begin{figure}
\centering
\epsfig{file=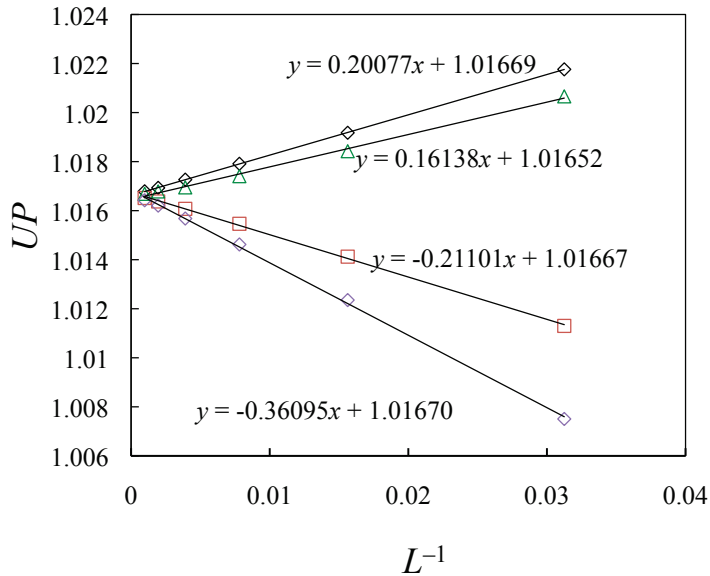,width=10cm}
\caption{Plots critical values of $y = UP$ for bond percolation (top), 
site percolation counting only occupied sites (second to top), 
the SIR model (second to bottom), and site percolation counting both 
occupied and surrounding vacant sites (bottom), as a function of 
$x = L^{-1}$.  The equations represent a linear fit to the simulation 
data; the error bars are smaller than the size of the symbols.}
\label{UPcriticalplot}
\end{figure}

In figure \ref{UPcriticalplot} we plot $UP$ at the critical point 
$c_c = 0.1765$ for systems of sizes $L = 32$, $64$, $128$, $256$, 
and $1024$.  Here we did at least $10^8$  samples for each size.  
As seen in that figure, we find a very good fit assuming the 
finite-size corrections are proportional to $1/L$.
The data extrapolate to a value of $UP = 1.0167(1)$ for an infinite system.

We also ran similar simulations of standard site and bond percolation 
at their critical points $p_c = 0.592746$ and $p_c = 0.5$ respectively, 
using an epidemic growth algorithm to generate the percolation clusters, 
and also show those results in figure \ref{UPcriticalplot}. 
For bond percolation, we characterized the cluster by the number of 
sites that are visited or ``wetted."  For site percolation, we consider 
two definitions of the cluster mass: the first is the number of occupied 
sites of the cluster, which conforms to the standard definition in 
percolation, while for the second we used both the number of occupied 
sites and the number of vacant sites surrounding the clusters (the 
so-called perimeter sites) to characterize the cluster size.  
The latter definition corresponds to the site-percolation limit of 
the SIR model, in which an I site simultaneously infects all its S 
neighbors with probability $p$, and then recovers, so that the R sites 
correspond to both occupied and vacant sites of the percolation cluster.  
 These systems show similar correction-to-scaling 
$\approx L^{-1}$, but with different coefficients---positive for bond 
and regular site percolation, negative for the SIR model and the 
occupied+vacant form of site percolation. 
The extrapolation of all three systems to $L \to \infty$ is a common 
value $1.0167(1)$, showing that the SIR model is equivalent to 
percolation not just for the critical exponents but for this amplitude 
ratio as well. 

Note that a correction-to-scaling behavior of $L^{-1}$ is often seen in percolation 
problems when there are boundaries or lattice effects present 
\cite{ziff92,hovi96}, but the precise source of the corrections to 
scaling is not clear here.   Also, it is not clear that the exponent is exactly
$-1$; the data in figure \ref{UPcriticalplot} is well fit for exponents
in the range of $-1$ to $-1.1$ (ignoring
higher-order corrections), depending upon the curve
.
While the universality of amplitude ratios in percolation has been 
studied for many combinations of quantities 
(e.g., \cite{aharony80,saleur85, adler86,privman91,aharony97,daboul00}), 
it seems that the quantity $UP$ has not been examined previously for percolation,
either away from, or exactly at, the critical point (finite-size
scaling).
For scaling away from $p_c$, $UP$ corresponds to the critical amplitude $R_3$ 
or $v_3$, which has been studied for the Ising model \cite{watson69,zinn96,pelissetto02}
but evidently not for percolation.  For finite-size scaling at the critical
point in percolation, studies have been carried out on other universal 
amplitude ratios \cite{aharony97,daboul00}, but not for $UP$.

The value of $UP$ is very close to $1$.  This can be understood easily for
standard percolation as follows: at the critical point, the largest cluster
$s_\mathrm{max}$ will be of the order of the size of the system
and much bigger than the other clusters, implying
$S \approx s_\mathrm{max}^2/L^2$, $M \approx s_\mathrm{max}^3/L^2$,
and (by definition) $P =  s_\mathrm{max}/L^2$, thus yielding $UP \approx 1$.
By universality, this also applies to the SIR model.
As can be seen in figure \ref{up_colapso}, for $c < c_c$ (which 
corresponds to $p > p_c$ for regular percolation), $UP$ goes to the
value exactly 1, as would be expected by these arguments.

\section{Conclusion}
\label{5}

We have used numerical simulations to investigate the critical behavior
of the stochastic SIR model on finite square lattices.
We have determined the order parameter $P$, the mean number
of recovered individuals $S$ and the mean squared number
of recovered individuals $M$. These quantities obey the
same scaling laws used in percolation theory.
The cluster size distribution of percolation theory is identified with
the cluster distribution of recovered individuals generated
by the SIR dynamics. By studying lattices of different sizes
we obtain the critical behavior by means of
a finite-size scaling borrowed from the stardard  percolation theory.
The value of the critical exponents are in agreement
with those of the isotropic percolation as one would expect.
We have shown that the ratio $U=M/S^2$ is not universal
at the critical point, as is the case of the models belonging
to the directed percolation universality class, and diverges
as $L^{\beta/\nu_\perp}$. Instead,
we have shown that the quantity that is universal is the product $UP$.
The value of the critical point $c_c$ found from the 
fact that $UP$ is independent of the system size is found
to be in excellent agreement with previous calculations.
The value of $UP$ at the critical point, $1.0167(1)$ is shown
to be consistent with measurements of standard site and bond percolation,
thus confirming that the amplitude ratios of the two models are the
same, and showing a deeper level of agreement between the SIR
and percolation models than just having common critical exponents.

\section*{Acknowledgment}

We wish to acknowledge the financial support of the
Brazilian agency CNPq, and RMZ acknowledges 
partial support from the U. S. National Science Foundation Grants 
No. DMS-0553487. 


\end{document}